\documentclass[doublecol]{epl2} 

\title{Modelling bacterial flagellar growth}
\shorttitle{Modelling bacterial flagellar growth} 

\author{M. Schmitt \and H. Stark}
\shortauthor{M. Schmitt \etal}

\institute{                    
  Institut f{\"u}r Theoretische Physik, Technische Universit{\"a}t Berlin - Hardenbergstra{\ss}e 36, 10623 Berlin, Germany, EU
}

\abstract{
The growth of bacterial flagellar filaments is a self-assembly process 
where flagellin molecules are transported through the narrow core of the 
flagellum and are added at the distal end. To model this situation,
we generalize a growth process based on the TASEP model by allowing 
particles to move both forward and backward on the lattice. 
The bias in the forward and backward jump rates determines the lattice 
tip speed, which we analyze and also compare to simulations.
For positive bias, the system is in a non-equilibrium steady state and
exhibits boundary-induced phase transitions. The tip speed is constant.
In the no-bias case we find that the length of the lattice grows as
$N(t)\propto\sqrt{t}$, whereas for negative drift $N(t)\propto\ln{t}$. 
The latter result agrees with experimental data of bacterial flagellar 
growth.}

\begin{document}

\maketitle

Bacterial flagella act as motility devices that allow bacteria such as 
\textit{Escherichia coli} and \textit{Salmonella} to swim and to respond 
to chemical stimuli by performing chemotaxis \cite{berg}. 
A flagellum mainly consists 
of a long helical hollow structure with a length of up to $20\mu m$ and a 
typical outer diameter of $0.02 \mu m$. The growth mechanism of the flagellar 
filament is a self-assembly process. Flagellin molecules are transported 
through the hollow core of the filament and are added one by one at the tip 
of the filament. An important quantity of such a growth process is the 
time dependence of the filament length $N(t)$. A recent model that treats 
flagellin molecules as diffusing particles in a single-file process 
established a growth function $N(t) \propto \sqrt{t}$ \cite{keener2006}. 
\revision{However, the only available experimental data to our knowledge show a logarithmic growth 
$N(t)\propto \ln t$ \cite{iino1974}.}
To model the growth process and to try to verify the experimental data, 
we make use of a variant of the well known totally asymmetric simple 
exclusion process (TASEP). This model takes into account the important fact 
that flagellar growth happens in a single-file process meaning 
flagellin molecules cannot pass each other.

The TASEP itself has developed into a paradigmatic 
model of non-equilibrium physics \cite{derrida1, derrida2, schuetz, krug, parmeggiani, mukamel}. 
In particular, TASEP models with open boundaries showed to be very fruitful 
for the study of non-equilibrium steady states, i.e., states that are 
characterized by non-vanishing currents. These systems have been used to 
model the motion of ribosome along mRNA \cite{macdonald}, molecular motors 
along microtubule filaments \cite{frey}, or that of cars on a highway 
\cite{barlovic, chowdhury}, just to name a few applications. All of these models 
consider lattices with a fixed number of lattice sites $N$. 
However, there are recent approaches that generalize the TASEP to 
dynamically extending exclusion processes. One approach considers the 
TASEP with a fluctuating boundary that can be pushed away by particles 
on the lattice \cite{nowak}, another one takes into account a death 
probability of the leading particle \cite{dorosz, kwon}.
In a different variant of the TASEP, particles that reach the end of the 
lattice can transform into new lattice sites resulting in a dynamically 
extending lattice \cite{sugden}. Initially, the extending TASEP was 
formulated to model fungal hyphal growth \cite{sugden2}. 
In the reminder 
of this paper we will generalize this model to a partially asymmetric model 
where particles can move both forward and backward on the lattice. In particular, 
we calculate the growth function $N(t)$ and discuss implications 
for the bacterial flagellar growth.
\begin{figure}
\centering
\includegraphics[trim = 0mm 210mm 0mm 0mm, clip, width=0.45\textwidth]{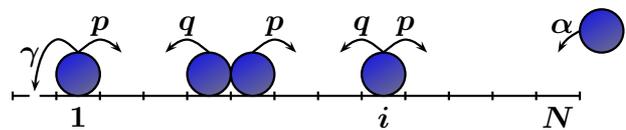}
\caption[Growing ASEP model]{Growing asymmetric simple exclusion process 
with open boundary conditions. The labels indicate rates at which particle 
hops can occur. At site $1$ particles transform into a new lattice site 
with rate $\gamma$, which then becomes the new site 1.}
\label{fig:1}
\end{figure}

The model is defined in the following way (see fig. 1 for a schematic). 
Particles enter the lattice with rate $\alpha$ at the boundary 
on the right. They either move to the left with rate $q$ or to the right 
with rate $p$ so that $q+p=1$. Finally, at the tip of the lattice at site 
$1$ they transform into a new lattice site with rate $\gamma$. A crucial 
feature of TASEP models is that particles can only move if the target 
site is empty. This hard-core interaction results in a nonlinear 
relationship between current $J$ and density $\rho$, which itself is the 
origin of the many interesting features of the TASEP.

\begin{figure}
\centering
\includegraphics[trim = 0mm 90mm 0mm 0mm, clip,width=0.45\textwidth]{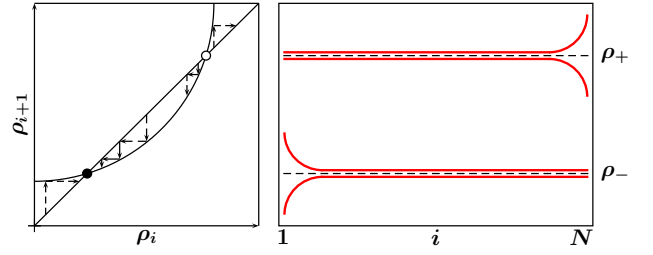}
\caption{Sketch of possible solutions to recurrence relation 
(\ref{growing_recurrence}) on the left and corresponding density profiles 
on the right. $\rho_-$ is a stable fixed point whereas $\rho_+$ is unstable.}
\label{fig:2}
\end{figure}

\begin{figure}
\centering
\includegraphics[width=0.45\textwidth]{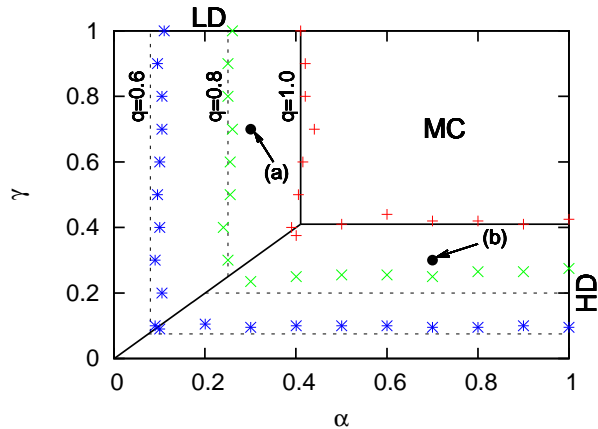}
\caption{Phase diagram of extending TASEP. Monte Carlo simulation 
(symbols) and improved mean-field calculations (lines) are shown for $q=1.0$, 0.8, and 0.6. 
Horizontal lines are boun\-da\-ries between HD and MC phases. Vertical 
lines separate LD from MC phases. Bullets: parameters of density plots in 
figs. \ref{fig:4}(a), (b).}
\label{fig:3}
\end{figure}

In a first step we study systems where particle hops possess a positive 
drift $q>p$. To begin the analysis, we write down the averaged 
rate equations for occupancy $ n_i\in\{0,1\}$ of site $i$:
\begin{eqnarray*}
\dot{\langle n_1\rangle}&=&q\langle(1- n_1) n_2\rangle -p\langle(1- n_2) 
n_1\rangle -\gamma \langle n_1\rangle\;\;,\\ 
\dot{\langle n_2\rangle}&=&q\langle(1- n_2) n_3\rangle -q\langle(1- n_1) 
n_2\rangle  -p\langle(1- n_3) n_2\rangle\\
&&+p\langle(1- n_2) n_1\rangle -\gamma \langle n_1 n_2\rangle\;,\\
\dot{\langle n_i\rangle}&=&q\langle(1- n_i) n_{i+1}\rangle -q 
\langle(1- n_{i-1}) n_i\rangle\\
&&-p\langle(1- n_{i+1}) n_i\rangle+p\langle(1- n_i) n_{i-1}
\rangle\qquad i\geq 3\;.\\
&&+\gamma \langle n_1( n_{i-1}- n_i)\rangle\;,
\end{eqnarray*}
The term $\gamma \langle n_1\rangle$ in the first equation describes the 
growth of the lattice. Further terms proportional to $\gamma$ occur since 
after each growth event the lattice sites are renumbered. 
\revision{This corresponds to working in the frame of reference of the tip 
so that the new site at the left boundary becomes site $1$.}
To tackle this system of coupled equations, we employ a 
mean-field approximation by setting $\langle n_{i} n_j\rangle=\langle n_{i}
\rangle\langle n_j\rangle=\rho_i\rho_j$, where $\rho_i$ is the density of 
site $i$. We now identify the particle current $J_{i,j}$ from site $i$ to 
its neighboring site $j$ as
\begin{eqnarray}
J_{1,0}&=&\gamma\rho_1\label{j1_mf}\;, \nonumber \\
J_{2,1}&=&q(1-\rho_1)\rho_2 -p(1-\rho_2)\rho_1\;, \label{j2_mf}\\
J_{i+1,i}&=&q(1-\rho_i)\rho_{i+1}-p(1-\rho_{i+1})\rho_i-\gamma\rho_1\rho_i\;,
\; i\geq2\;.\label{j3_mf} \nonumber
\end{eqnarray}
In a steady state, the current $J$ is constant throughout the lattice. 
Moreover, since particles at the boundary on the left are used solely for 
the growth of the lattice, the current must exactly match the tip speed 
$J=v=\gamma\rho_1$. This allows us to solve the set of equations (\ref{j2_mf}) 
for the density throughout the lattice
\begin{equation}
\left. \begin{array}{rl}
\rho_1=\frac{v}{\gamma},\qquad
\rho_2=\frac{v(1+\frac{p}{\gamma})}{q-\frac{v}{\gamma}(q-p)},\qquad\\
\\
\rho_{i+1}=\frac{v + (v+p)\rho_i}{q-(q-p)\rho_i} \;,\qquad\quad
i\geq 2\; .\end{array} \right.
\label{growing_recurrence}
\end{equation}
The last equation shows two fixed points (see fig. \ref{fig:2}) upon setting 
$\rho_i=\rho_{i+1}$:
\begin{equation}
\rho_\pm=\frac{q-p-v}{2(q-p)}\pm\sqrt{\left(\frac{q-p-v}{2(q-p)}\right)^2-\frac{v}{q-p}}\; ,\label{rho_pm_growing}
\end{equation}
where $\rho_-$ is stable and $\rho_+$ is unstable as indicated in 
fig.\ \ref{fig:2}(a). Iterating eq.\ (\ref{growing_recurrence}) results in 
four possible density profiles [see fig.\ \ref{fig:2}(b)]. Whereas the
low-density profiles relax towards the stable fixed point $\rho_-$, the 
high-density profiles start in close vicinity of the unstable fixed point
$\rho_+$ and deviate from it when the other end of the lattice is reached.
Upon changing bias $q$ and boundary conditions $\alpha, \gamma$, it is now
possible to identify three different phases. For low input rate $\alpha$ at 
lattice site $N$, the system is in the low density (LD) phase, whereas low
growth or release rate $\gamma$ at lattice site 1 leads to the high 
density (HD) phase. The crossover between low density and high density 
phase is governed by a first-order transition, where both phases coexist. 
The locations of the phases as determined by the mean-field theory are similar to the refined
$\alpha,\gamma$ phase diagram in fig. \ref{fig:3}, which we will discuss below.
When both rates $\alpha$ and $\gamma$ are high, the fixed points of 
eq.\ (\ref{fig:2}) do no longer exist and the system is in the maximal current
(MC) phase. The density profile is then no longer controlled by the boundary conditions
$\alpha$ and $\gamma$. Instead, the bottleneck in the system is given by the
bulk dynamics, namely due to the implicit hard-core interaction, which
restricts the current or tip speed to its maximum value $J_{MC}=v_{MC} =
(3-2\sqrt{2})(q-p)$. It is determined from eq.\ (\ref{rho_pm_growing}) by
setting the root to zero, i.e., when the two fixed points merge. Decreasing 
the hopping rate bias $q$ below $1$ slows down the bulk dynamics which results 
in an expanding maximal current phase in the phase diagram. 

In order to determine the phases just discussed within the formulated
mean-field theory, we extended and generalized the procedure of
ref. \cite{sugden} to $q<1$. We shortly summarize the reasoning.
Setting $\rho_{N-1}=\rho_N$ in the mean-field rate equation for the boundary
on the right,
\begin{equation}
 \dot{\rho_N}=p(1-\rho_N)\rho_{N-1}-q(1-\rho_{N-1})\rho_N+\alpha(1-\rho_N)\;,
\label{rho_N}
\end{equation}
yields the boundary condition $\rho_N=\alpha/(q-p)$. In the LD phase, 
$\rho_N$ equals the stable
fixed point $\rho_-$ of eq.\ (\ref{rho_pm_growing}) which gives an expression 
for the LD current or tip speed $J_{LD}=v_{LD}$. However, the iteration of
eq.\ (\ref{growing_recurrence}) starting at $\rho_2$ only converges to 
the stable 
fixed point $\rho_-$ if it starts below the unstable fixed point $\rho_+$,
i.e., if $\rho_2 < \rho_+$. Using the expression for $J_{LD}=v_{LD}$, this
condition leads to a phase boundary against the HD phase. In the HD phase the
tip velocity $v_{HD}$ and the high density value $\rho_+$ follow by setting 
$\rho_2=\rho_+$. From $\rho_2=\rho_+$ the density then has to decay towards 
$\rho_N=\alpha/(q-p)$ at the boundary on the right. This sets up a 
stationary shock profile at the phase transition between LD and HD phase. 
Close to this phase transition in the HD phase, the shock moves away from 
the tip but never reaches the other boundary of the lattice. Only further 
away from the phase transition, does the high density expand throughout 
the lattice except in a narrow region close to the boundary on the right.

\begin{figure}
\centering
\includegraphics[trim = 0mm 0mm 27mm 19mm, clip, width=0.48\textwidth]{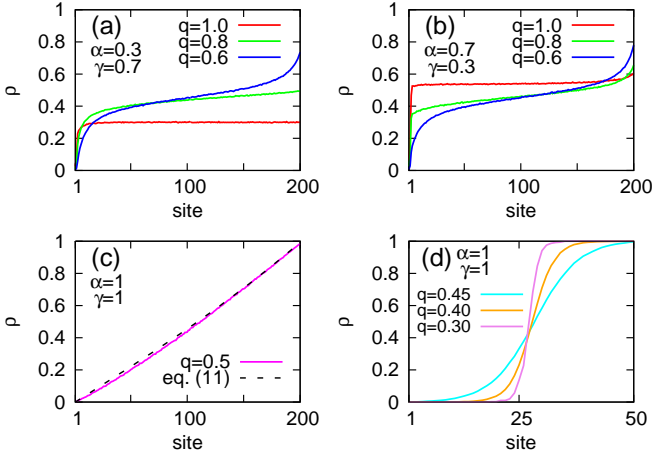}
\caption{
Monte Carlo simulations of density profiles for different phases:
(a) LD phase with crossover to MC phase ($q=0.8, 0.6$), (b) HD phase with 
crossover to MC phase ($q=0.8, 0.6$) [Bullets in fig. \ref{fig:3} give the
location in the phase diagram], (c) $q=p$, and (d) $q<p$.}
\label{fig:4}
\end{figure}

\revision{Monte Carlo simulations of our model were done using a random sequential update where sites 
are picked randomly one after another and updated according to rates $\alpha, \gamma, q$. 
One time step consists of updating all $N$ lattice sites. The tip velocity $v$ is then determined as a function
of either $\alpha$ or $\gamma$ for a given value of $q$. We localize the HD-MC or LD-MC phase
transition when $v$ reaches a constant maximum value which indicates the MC 
phase.}
As already indicated, the resulting phase diagram obtained by the mean-field 
approach does not coincide very well with results from Monte-Carlo
simulations. An improved mean-field calculation that takes into account 
correlations between site $1$ and $2$  by setting 
$\langle n_1 n_2 n_i\rangle=\langle n_1 n_2\rangle n_i$, 
predicts a more precise phase diagram. This was already observed for $q=1$,
i.e., in the totally asymmetric case \cite{sugden}. In contrast to the
mean-field approach discussed above, iteration of eq.\
(\ref{growing_recurrence}) then starts at site $3$ instead of site $2$. In 
particular, to eliminate $\langle n_1 n_2\rangle$ in the equation for $\rho_3$,
one needs to evaluate the rate equation for the state 
$\langle n_2(1-n_1)\rangle$. 
The resulting phase diagram is shown in fig. \ref{fig:3} and compared to 
Monte-Carlo simulations. The improved mean-field calculations correctly predict
the phase boundaries between LD and MC phases for various $q$ values whereas
the boundaries between HD and MC phases still show some deviations from simulations.
\revision{This was already observed in \cite{sugden}. Since determining the HD-MC boundary involves 
the starting value of iteration (\ref{growing_recurrence}), the result depends strongly 
on the number of site correlations that are taken into account at the tip.}
Figures\ \ref{fig:4}(a) and (b) show density profiles at constant 
$\alpha$, $\gamma$ that convert from the respective LD or HD phase at $q=1$
to the MC phase at $q=0.8$ and 0.6.

For our purposes the main conclusion is that the system reaches a
non-equilibrium steady state characterized by a constant current or tip 
speed $v$. Hence, the lattice grows linearly in time: $N(t)=v t$. This 
is valid for all three phases and the actual value of $v$ depends on
the parameters $\alpha$, $\gamma$ and $q$. 
With $v_{MC}=(3-2\sqrt{2})(q-p)$ in the MC phase, we obtain
\begin{equation}
N(t)=(3-2\sqrt{2})(q-p)t\;,\qquad q>p\;,\label{N_const}
\end{equation}
which agrees with simulations [see fig. \ref{fig:6}(a)]. Note that 
Eq.\ (\ref{N_const}) predicts $N(t)\rightarrow 0$ for decreasing bias 
$q\rightarrow 0.5$. However, Monte-Carlo simulations show a non-vanishing 
growth of the lattice for $q=p$. So, we have to study this case separately.

When $q=p=0.5$, the bulk of the lattice is not driven anymore, i.e., 
the condition of detailed balance is fulfilled in the bulk but not at the boundaries. 
The system is still kept out of equilibrium. For $q=p$, eqs.\ (\ref{growing_recurrence}) reduce to
\begin{equation}
\left. \begin{array}{rl}
\rho_1={v}/{\gamma},\qquad
\rho_2=2v+v/\gamma,\\
\\
\rho_{i+1}=\rho_i(1+2v)+2v\;, \quad i\geq 2\;.
\end{array} \right.
\label{growing_recurrence_2}
\end{equation}
The recurrence relation does not have a fixed point with $\rho > 0$
and the density grows to infinity, 
$\lim_{i\rightarrow\infty}\limits\rho_i=\infty$. To bypass this behavior,
we impose that the density reaches the physical maximum at the boundary
on the right, $\rho_N=1$. 
\revision{For $N \gg 1$ and with $\rho_N=1$ one readily shows that $\rho_1=v/\gamma\ll 1$ 
and therefore negligible in eqs. (\ref{growing_recurrence_2}). Then, the recurrence relation of (\ref{growing_recurrence_2}) is solved by}
\begin{equation}
 \rho_i=(1+2v)^{i-1}-1\;.\label{rho_i}
\end{equation}
The condition $\rho_N=1$ gives $2=(1+2v)^{N(t)}$, which indicates 
that $v$ is time-dependent now. By identifying $v(t)$ as ${dN(t)}/{dt}$,
we can solve for $N(t)$ and arrive at
\begin{equation}
 N(t)=\sqrt{\ln{2}}\sqrt{t} \qquad \mathrm{for} \qquad t \gg 1 \;. 
\label{N_sqrt_t}
\end{equation}
This result is confirmed by simulations as indicated in fig. \ref{fig:6}(a). 
It is independent of $\alpha$ and $\gamma$, i.e., the MC phase fills the 
whole phase diagram. It also shows that the current in the lattice is
ohmic, $J=v={dN(t)}/{dt} \propto 1/N(t)$.

We note that the growth function of eq. (\ref{N_sqrt_t}) is consistent 
with a continuum  approach for this particular system. The non-extending 
TASEP with fixed $N$, open boundaries, and $q=p$ reduces to the diffusion 
equation in the continuum limit. Hence, to obtain a continuous model of 
the extending TASEP with $q=p$, one has to formulate the diffusion equation 
on a growing domain. Similar to writing a reaction-diffusion equation
on a growing domain \cite{crampin1999}, we find \cite{schmitt2010}
\begin{equation}
\frac{\partial\rho}{\partial t}=\frac{1}{2 N(t)^2}
\frac{\partial^2\rho}{\partial x^2}-\frac{1}{N(t)}\frac{dN(t)}{dt}\rho,
\quad\quad x\in[0,1]\;,\label{edwards_wilkinson}
\end{equation}
where we replaced $\rho_i(t)$ by $\rho(x,t)$ and the growing domain
is mapped on the interval $[0,1]$. This equation only has a stationary
solution in the case of diffusive growth $N(t)\propto\sqrt{t}$. For
$\partial \rho/\partial t = 0$ and using growth function (\ref{N_sqrt_t}), 
we arrive at the time-independent equation
\begin{equation}
 \frac{1}{\ln 2}\frac{d^2\rho}{dx^2}-\rho=0\;.
\end{equation}
Under the assumption of Dirichlet boundary conditions $\rho(0)=0$ and 
$\rho(1)=1$, where the tip of the lattice is located at $x=0$, the 
stationary solution reads
\begin{equation}
 \rho(x)=\frac{\sinh (\sqrt{\ln 2}\, x)}{\sinh \sqrt{\ln 2} }\;.\label{sinh}
\end{equation}
 Monte Carlo simulations confirm this density profile (see fig. \ref{fig:4}(c)). 
The deviation from a linear density profile, i.e., the solution to the 
non-extending TASEP with $q=p$, is rather small. 

\begin{figure}
\centering
\includegraphics[trim = 0mm 200mm 20mm 0mm, clip, width=0.35\textwidth]{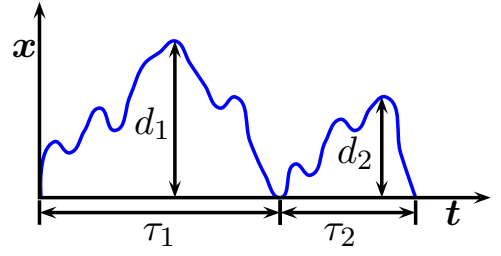}
\caption{Trajectory of a biased random walker with fixed wall at the origin. The random walker starts at $x=0$. The height of an excursion is called $d_i$, the length is called $\tau_i$.}
\label{fig:5}
\end{figure}

When $q<p$, the particles in the bulk prefer to hop to the right, even 
though the boundary conditions enforce a current from right to left. As a 
result, the growth function (\ref{N_const}) becomes negative indicating that 
the mean-field calculation is not applicable in this case. 
\revision{Studies of the non-extending TASEP with $q<p$ exist \cite{blythe, gier} and exact solutions 
revealed a reverse bias phase where the particle current decreases exponentially with system size $N$.}
In our case, we 
expect that the lattice grows very slowly for $q<p$. Furthermore, 
most of the particles will stay in a domain at the right boundary, 
whereas near the left boundary most of the sites are vacant. Monte 
Carlo simulations confirm that there is a domain interface between an 
empty domain and a "jammed" domain. As fig. \ref{fig:4}(d) shows, this 
domain wall is roughly located in the middle of the lattice. Now, 
whenever the left-most particle in the lattice reaches the tip, the 
lattice will extend by one site. After such a growth event, the next 
particle in the lattice becomes the left-most particle. 
Therefore, the growth of the lattice is equivalent to finding the
probability that a biased random walker reaches a maximum distance 
from a bounding wall. In this picture, the wall is given by the domain 
interface between the empty domain on the left and the "jammed" domain on 
the right and the maximum distance is the distance of the lattice tip from
the domain interface. Figure \ref{fig:5} depicts a trajectory of a 
biased random walker with fixed wall at the origin. The random walker is 
on excursions that last for a timespan of $\tau_i$ and have a maximum 
distance $d_i$ from the origin. Ref. \cite{iglehart} determines the
probability for $d_i$ to be smaller than some threshold $c$ in the
long-time limit as 
\begin{equation}
 P(d_i<c)=1-e^{-c/\kappa}\; , \label{distribution}
\end{equation}
where $\kappa$ characterizes the step-length distribution $W(X_i)$ of the
random walker by the condition
\begin{equation}
\langle e^{X_i/\kappa}\rangle = 1 \;.\label{kappa_assumption}
\end{equation}
In our case, the respective probabilities for hops to the left or right, 
i.e., away from or towards the origin, are $W(X_i=+1) = q $ and $W(X_i=-1)=p$. 
The condition (\ref{kappa_assumption}) gives
\begin{eqnarray*}
qe^{1/\kappa}+pe^{-1/\kappa}=1\;,
\end{eqnarray*}
which results in
\begin{equation}
\kappa=\left(\ln \frac{p}{q}\right)^{-1}\; .\label{kappa}
\end{equation}
At time $t$ the random walker made $t/\langle \tau_i\rangle$ excursions 
and the maximum $M_t$ is given by
\begin{equation}
 M_t=\mathrm{max}(d_1,d_2,\dots,d_{{t}/{\langle \tau_i\rangle}})\;.
\end{equation}
The probability that $M_t<c$ becomes
\begin{eqnarray*}
 P(M_t<c)&=&P(d_1<c)\cdot P(d_2<c)\dots 
P(d_{t/\langle \tau_i\rangle}<c)\\
&=&\left(1-e^{-c/\kappa}\right)^{t/\langle \tau_i\rangle}\;.
\end{eqnarray*}
Replacing $c$ by $c+\kappa\ln t$ yields
\begin{equation}
  P(M_t<c+\kappa\ln t)=\left(1-
\frac{e^{-c/\kappa}}{t}\right)^{{t}/{\langle \tau_i\rangle}}\;
\end{equation}
and in the limit of large $t / \langle \tau_i\rangle$, one obtains the 
Gumbel distribution \cite{gumbel}
\begin{equation}
  P(M_t<c+\kappa\ln t)=\exp(-e^{-c/\kappa}/\langle \tau_i\rangle)\;.
\end{equation}
This means that the probability for a maximum excursion stays constant
in time, when in the long-time limit the maximum grows as
\begin{equation}
 M_t = \kappa \ln t + c\;,  
\end{equation}
where $c$ is an undetermined constant. Since the domain interface is 
situated approximately in the middle of 
the lattice, the growth function of the whole lattice is then
\begin{equation}
 N(t) = 2\left(\ln \frac{p}{q}\right)^{-1} \ln t + c\;,\qquad\qquad q<p\;,
\label{N_lnt}
\end{equation}
\revision{
with tip speed
\begin{equation}
 v=\frac{dN(t)}{dt}\propto \exp(-\frac{N}{2\kappa})\;.
\end{equation}
Interestingly, the dependence of $v$ on $N$ agrees qualitatively with the dependence of the current on $N$ in 
the non-extending TASEP with negative drift \cite{blythe, gier}. 
}
\begin{figure}
\centering
\includegraphics[trim = 0mm 0mm 43mm 41mm, clip, width=0.49\textwidth]{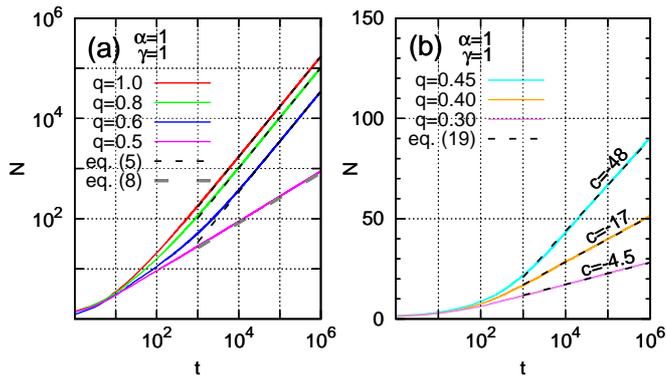}
\caption{Lattice growth determined from Monte Carlo simulations. 
(a) Upper 3 curves: positive drift with $q>p$, bottom curve: no drift 
with $q=p=0.5$. (b) Negative drift with $q<p$.}
\label{fig:6}
\end{figure}
Simulations of an extending lattice for different $q < 0.5$ agree very 
well with this theory by reproducing the logarithmic growth law 
(\ref{N_lnt}) with its prefactor $2\kappa$ [see fig. \ref{fig:6}(b)].
This is even true for $q$ approaching 0.5 when the domain interface in
fig. \ref{fig:4}(d) becomes broader and, strictly speaking, the bounding 
wall of our model does not exhibit a hard-core repulsion for the 
random walker. Just as in the $q=p$ case, simulations of an extending lattice are independent of $\alpha$ and $\gamma$. 
Note that the analogy with a biased random walker bounded
by a wall also gives the growth laws $N(t)\propto t$ for $q>p$ and 
$N(t)\propto \sqrt{t}$ for $q=p$. We checked this by Monte-Carlo
simulations.
\begin{figure}
\centering
\includegraphics[width=0.45\textwidth]{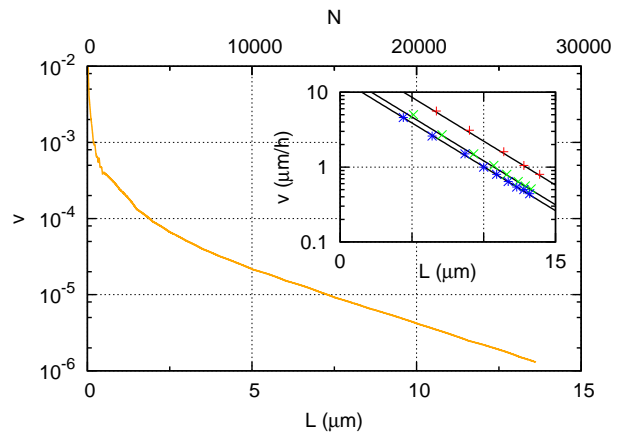}
\caption{Lattice growth speed from simulations for small negative 
bias ($q=0.4999$). \revision{The inset shows flagellar elongation data of three samples of Salmonella and fits, taken from \cite{iino1974}.}}
\label{fig:7}
\end{figure}

Summarizing our results, we found three distinct regimes for the bias $q$ 
in the extending TASEP with qualitatively different growth functions. The 
experimental observation of logarithmic growth for flagellar filaments 
\cite{iino1974} mentioned in the introduction is in qualitative agreement 
with our model when the drift is negative ($q<0.5$). In order to compare 
our result with the experiment in more detail, we carried out simulations 
with very small negative drift. By setting $q=0.4999$, it was possible to 
achieve a lattice length of the order of $N=20000$ in reasonable simulation 
time while still reaching the logarithmic growth regime. If each growth 
event mimics one flagellin molecule, $N=20000$ corresponds to a filament 
of about $10\mu m$. For our estimate, we used the length $55 \mathring{A}$ for a flagellin 
molecule and the fact that 11 flagellin molecules make up the circular 
cross section of the hollow flagellar filament. The speed of elongation 
of flagellar filaments in the experiment was found to decrease 
exponentially with increasing length for filament lengths in the range 
from $4$ to $14 \mathrm{\mu m}$ \revision{(see inset of fig. \ref{fig:7})} \cite{iino1974}. 
This is in agreement with our simulations \revision{as illustrated in fig. \ref{fig:7}.} 
However, it remains to establish an explanation why the flagellin in the hollow 
filament should exhibit a negative drift towards the cell body. It has 
been suggested that proton motive force (PMF) is biasing the Brownian 
motion of the flagellin translocation process \cite{minamino}. The 
details of this mechanism are, however, very unclear at this stage 
and should be subject to further research.

\end{document}